\documentclass[11pt]{article}

\usepackage{amsmath}
\usepackage{amsfonts}
\usepackage{amssymb}
\usepackage{lscape}
\usepackage[T1]{fontenc}
\usepackage[latin1]{inputenc}
\usepackage{geometry}
\usepackage{comment} 
\begin{document}

\title{Real-time and Probabilistic Temporal Logics: An Overview}

\author{Savas Konur\\Department of Computer Science, University of Liverpool}

\date{}

\maketitle

\begin{abstract}
Over the last two decades, there has been an extensive study on logical formalisms for specifying and verifying real-time systems. Temporal logics have been an important research subject within this direction. Although numerous logics have been introduced for the formal specification of real-time and complex systems, an up to date comprehensive analysis of these logics does not exist in the literature. In this paper we analyse real-time and probabilistic temporal logics which have been widely used in this field. We extrapolate the notions of decidability, axiomatizability, expressiveness, model checking, etc. for each logic analysed. We also provide a comparison of features of the temporal logics discussed.
\end{abstract}

\section{Introduction}

Temporal logics have been extensively used in the specification of various systems, such as real-time and control systems, for more than two decades. They provide a mathematical foundation to formally analyse these systems. Many industrial applications and case studies proved the usability of temporal logics within this context. 

A system behaviour is usually described by a set of `events', and their associated `temporal constraints'.  Temporal logics allow us to express such a behaviour by means of `logical formulas' \cite{BMN00}. In general, temporal logics have been introduced for specific types of problems. The general trade-off is between the complexity and simplicity. In certain applications simple logics are preferred to the complex ones \cite{BMN00}. Complex logics are generally difficult to deal with practically. 

Numerous logics have been introduced for the formal specification of real-time and complex systems, and various aspects of logics have been studied. Some surveys \cite{Ost92,AHT92,BMN00,GMS04} make a comprehensive analysis of specific logics. In this paper we outline main and recent developments in the field in a broad sense. Namely, we give an overview on most-known temporal logics introduced up to now. All these logics are different in terms of `expressiveness', `order', `time metric', `temporal modalities', `time model' and `time structure'. They also have different capabilities for the specification and verification of real-time systems.

In this paper we survey the following aspects: `basic temporal framework', `real-time' and `probability'. Real-time aspect of temporal logics is important to express timing requirements of real-time systems. Probabilistic aspect is needed in order to reason about systems which include uncertainty and probabilistic assumptions. 

In the following we will analyse well-known real-time and probabilistic temporal logics. We will summarize important results on decidability, axiomatizability, expressiveness, model checking, etc. for each logic analysed. We will also provide a comparison of features of the temporal logics discussed.  

Note that in some instances we think it is more convenient to refer to the original text for clarification purposes. In the following,  we will use quotation marks to use the text from the original resources.

\section[Pre]{Preliminaries\footnote{This section is taken verbatim from \cite{Kon10}.}} \label{sec:Classification-of-Temporal}

We can classify temporal logics based on several criteria. The common dimensions are `propositional versus first-order', `point-based versus interval-based', `linear versus branching', `discrete versus continuous', etc \cite{Eme95,Ven98,BMN00}. Below we discuss the most important  criteria to classify temporal logics.

\paragraph{Point versus interval structures:} There are two structure types to model time in a  temporal logic: \emph{points (instants)} and \emph{intervals}. A point structure $\mathcal{T}$  can be represented as $\left\langle T,<\right\rangle$, where $T$ is a nonempty time points, and $<$ is a `precedence' relation on $T$. Different temporal relationships can be described using different modal operators. Some logics include modal operators  which can express quantification over time. However, a relationship between intervals is difficult to express using a point-based temporal logic \cite{FM94}. 

Interval temporal logics are expressive, since these logics can express a relationship between two events, which are represented by intervals. Also, interval logics \cite{SMS82,SMV83,Mos83,Lad87,MS87,RG89,HS91} have a simpler and neater syntax to define a relationship between intervals, which provides a higher level abstraction than a point-based logic when modeling a system. This makes interval logic formulas much simpler and more comprehensive than point-based logic formulas. 

Some of the known interval operators are \emph{meets}, \emph{before}, \emph{during} \cite{All83}, which denote the ordering of intervals; \emph{chop} modality \cite{Ven91}, which denotes combining two intervals; and \emph{duration}, which denotes a length of an interval \cite{CH04}.

Interval structures can be considered in two ways: \emph{(i)} intervals are `primitive' objects \emph{(ii)} intervals are composed from points. \cite{vB91,MSV02,Vit05} consider intervals as primitive objects of time. \cite{vB91} defines  a `period structure' as the tuple $\left\langle \mathcal{I},\subseteq,\prec\right\rangle $, where $\mathcal{I}$ is a non-empty set of intervals, $\subseteq$ is a sub-interval relation, and $\prec$ is a precedence relation.
One particular problem of this approach is that theoretical analyses are usually very difficult. Also, although it is very easy to define properties  \emph{linearity}, \emph{density}, \emph{discreteness}, \emph{unboundedness} in a point-based logic, it is very difficult to define these properties in an interval logic where intervals are primitive objects. 

\cite{GMS04,HS91,Ven91} consider intervals as set of points, where the time flow is assumed as ``a strict partial ordering of time points''. Namely, an interval structure is defined as $\left\langle \mathcal{T},\mathcal{I}(\mathcal{T})\right\rangle$, where $\mathcal{T}=\left\langle T,<\right\rangle $ is a strict partial ordering and $\mathcal{I}(\mathcal{T})$ is a set of intervals. The properties mentioned above can be defined in an interval logic where intervals are composed of time instants. 

We conclude this section with the historical development of interval-based temporal logics. The concept of time intervals was first studied by Walker \cite{Wal47}. Walker considered a non-empty set of intervals, which is partially orderd. However, his work does not cover aspects of temporal logic in a general sense. In \cite{Ham71}  philosophical aspects of an interval ontology was analysed. In \cite{Hum79} an interval tense logic was introduced. 
 \cite{Dow79,Kam79,Rop80,Bur82,vB83,Gal84,Sim87} studied interval logics within the natural language domain. 
 It was argued that interval-based semantics are more convenient for human language and reasoning, and 
interval-based approach is more suitable than point-based approach for temporal constructions of natural language.
\cite{All83,All84,AH89,AF94} studied event relations and interval ordering. The authors introduced so-called Allen's thirteen interval relations and worked on axiomatisation and representation of interval structures.  Some further works on Allen's algebra were carried out by \cite{Lad87,Gal90}. Recently, \cite{RB06} investigated the relation between Allen's logic and LTL. Interval based-logics have been also applied to other fields in computer science. \cite{Par78,Pra79,HPS83} worked on process logic, where intervals are used as representation of information. Another important work was the development of \emph{interval temporal logic} \emph{(ITL)}, and its application to design of hardware components \cite{Mos83,HMM83}. Since the development of  ITL, various variations have been proposed so far. In particular, Duration Calculus \cite{CH04} is an extension of interval temporal logic with ``a calculus to specify and reason about properties of state durations''.

\paragraph{Temporal Structure:} There are important properties regarding the time flow and temporal domain structure. Some properties are summarized below:

Assume  $\left\langle T,<\right\rangle$ represents a temporal structure, where $T$ is a nonempty time points, and $<$ is a `precedence' relation on $T$. In a temporal logic the structure of time is \emph{linear} if any two points can be compared. Mathematically, a strict partial ordering is called linear if any two distinct points satisfy the condition: $\forall x,y:x<y\vee x=y\vee x>y$. This definition suggests that in linear temporal logics each time point is followed by only one successor point. 


Another class is the \emph{branching-time structures}, where the underlying temporal structure is  branching-like, and each point may have more than one successor points. The structure of time can be considered as a tree. A \emph{tree} is a set of time points $T$ ordered by a binary relation $<$ which satisfies the following requirements \cite{GHR94}:

\begin{itemize}
\item \noindent $\left\langle T,<\right\rangle $ is irreflexive;
\item \noindent $\left\langle T,<\right\rangle $ is transitive;
\item \noindent $\forall t,u,v\in T$ $u<t$ and $v<t \rightarrow$
$u<v, u=v$ or $u>v$ (i.e. the past of any point is linear);
\item \noindent $\forall x,y\in T, \exists z\in T$ such that $z<x$
and $z<y$ (i.e. $\left\langle T,<\right\rangle $ is connected).
\end{itemize}

One important characteristics of branching logics is that the syntax of these logics include path quantification which allows formulas to be evaluated over paths. However, linear temporal logics are restricted to only one path.

A temporal domain is \emph{discrete} with respect to the precedence relation $<$ if each non-final point is followed by a successor point. This can be formulated as follows: $\forall x,y$ ($x<y \rightarrow \exists z$ ($x<z \wedge \neg\exists w (x<w \wedge w<z$))) \cite{Spr02}. Majority of temporal logics used for system specification are defined on discrete time, where points represent system states. A state sequence, as a result of a program execution, can be considered as isomorphic to discrete series of positive integers. 

A temporal domain is \emph{dense} if, between any two distinct points, there is another point. This can be formally denoted  $\forall x,y (x<y \rightarrow \exists z (x<z<y$)) \cite{Spr02}. Above we mentioned that flow of discrete time can be represented as positive integers. Similarly, density can be represented as real numbers. It is noteworthy to mention that there is a distinction between \emph{density} and \emph{continuity}: ``A model of dense time is isomorphic to a dense series of rational numbers, meaning that there is always a rational number between any two rational numbers; whereas a model of \emph{continuous} time is isomorphic to a continuous series of real numbers'' \cite{Ven98}.

A temporal domain is \emph{bounded above} (\emph{bounded below}) if the temporal domain is bounded in the future (past) time. This can be formulated as follows: $\exists x \neg\exists y (x<y$ ($\exists x \neg\exists y (y<x$))) \cite{Spr02}. Similarly, a temporal domain is \emph{unbounded above} (\emph{unbounded below}) if each point has a successor (predecessor) point, which is formally denoted  $\forall x \exists y (x<y$  ($\forall x \exists y (y<x$))) \cite{Spr02}. 

A temporal domain \emph{is Dedekind complete} if all time point sets (non-empty) are bounded above, and they have a least upper bound.

Based on differences in temporal domain properties logics have different characteristics. For example, we can consider a temporal domain which is linear or branched; discrete or dense; finite/infinite in future and/or past, etc. All these choices result in different syntax, semantics, decidability and complexity. 

\section{Real-Time Temporal Logics}

Over the last two decades, temporal logics have been used as a mathematical foundation to formally analyse real-time systems. System behaviours are usually expressed in terms of a logical formula. Although this depends on the richness of the language, in general, temporal logics are very expressive to specify important aspects of the systems. Generally speaking real-time temporal logics have been defined for specific purposes. In certain cases, temporal logics with a simple syntax are used in order to make them practically feasible. 

Below we give a brief account of well-known real-time temporal logics (summarised from \cite{Ost92,AHT92,BMN00}). All these logics are different in terms of `expressiveness', `order', `time metric', `temporal modalities', `time model' and `time structure'. They also have different capabilities for the specification and verification of real-time systems.

\subsection{Real-time Extensions of CTL}

In \cite{EMSS89} a real-time extension of CTL, called RTCTL, was introduced. RTCTL has ``point-based strictly-monotonic integer-time semantics'' \cite{AHT92}. RTCTL includes a metric for time. The satisfiability problem of RTCTL is 2-EXPTIME-complete \cite{EMSS89}. The model-checking problem is linear \cite{EMSS89}.

\cite{ACD90} introduced the real-time logic TCTL, which extends CTL with hidden clock bounded operators. It has ``point-based strictly-monotonic real-time semantics'' \cite{AHT92}.  The satisfiability checking of a TCTL formula is undecidable if it is interpreted over dense time domains; but the model checking problem still remains decidable \cite{ACD90}. \cite{ACD90} finds that the model checking complexity of TCTL is ``exponential in the number of clocks, exponential in the length of timing constraints, linear in the size of the node-transition graph, linear in the number of operators in the formula and exponential in the length of the subscripts in the formula''. \cite{ACD90} also shows that the upper bound can be improved to PSPACE, and the model checking problem is PSPACE-complete. \cite{LMS04} considers the model checking problem of different subclasses of TCTL. 

Another branching time logic called TPCTL is introduced in \cite{Han91}. TPCTL is a probabilistic extension of CTL. The underlying time structure is represented by discrete time. TPCTL semantics is defined over non-deterministic probabilistic transition systems. TPCTL can express both hard and soft deadline properties, such as
`an error occur with a probability of 0.1 within 100 seconds'.  \cite{Han91} shows that TPCTL model checking has EXPTIME complexity. \cite{BS98} proves that the model checking problem is polynomial.

\subsection{Real-Time Logic (RTL)\label{sub:Real-Time-Logic-(RTL):}: }

RTL is a first-order temporal logic, introduced in \cite{JM86} to reason about events and their relations. The logic includes a so-called \emph{occurrence} function which maps each event to a time stamp. Existence of occurrence function allows RTL to express periodic and non-periodic real-time properties.

In RTL, time is measured with an `absolute' clock whose value can be referenced in a formula.  RTL is defined over a linear sequence of discrete time points, which are bounded in the past, but unbounded in the future. \cite{AH90} shows that under these semantics RTL is undecidable. 

Since absolute clocks are used, and clock values can be explicitly referenced in formulas, RTL can be used to express ordering and quantitative temporal constraints.  One disadvantage of this functionality is that using explicit reference to time results in complex formulas difficult to understand.  For example, the temporal constraint ``for each occurrence of an event \emph{B} which happens at a time instant \emph{$t_{0}$,} the predicates \emph{startA} and \emph{endA} hold (marking an interval {[}\emph{startA}, \emph{endA}] at which \emph{A} is true), and the interval {[}\emph{startA}, \emph{endA}] is subsumed by the interval {[}\emph{$t_{0}$}, \emph{$t_{0}$+$t_{b}$}] (where \emph{$t_{0}$}$\leq$\emph{startA} $\leq$ \emph{endA}$\leq$\emph{$t_{0}$+$t_{b}$})''  is specified in RTL as follows \cite{BMN00}:

\begin{itemize}
\item $\forall t.\forall i.@(\Omega B,i)=t \rightarrow(\exists j.(t \leq @(\uparrow A,j))\wedge(@(\downarrow A,j)\leq t+t_{b}))$
\end{itemize}

\noindent where $\Omega B$ denotes the occurrence of the event \emph{B}, \emph{t} denotes time, \emph{$\uparrow A$} denotes the beginning of the action \emph{A}, \emph{$\downarrow A$} denotes the completion of the action \emph{A}, and \emph{i} and \emph{j} are the occurrences of the events marked with the operator $@$. Time is captured by the occurrence function $@$ which assigns time values to event occurrences; $@(\Omega B,i)$ is defined as the time of the \emph{i}-th occurrence of $\Omega B$ \cite{Ost92}.

Decision procedures devised for RTL in general are not practical. To increase the efficiency some methods were deployed. In \cite{JS88}, RTL formulas are re-structured into ``computational graphs'' using a formalism called ``modecharts'',  which resulted in ``an exponential time decision procedure (in the worst case)'' \cite{Ost92}. 

\subsection{Real-Time Temporal Logic (RTTL):}

RTTL \cite{OW87,Ost89} is a first-order explicit clock logic. Discrete linear time points are employed as temporal structure. The sequence of time points are bounded in the past, but unlimited in the future. In an RTTL formula the clock variable  \emph{t} is explicitly referred. RTTL is a first-order logic because any arbitrary quantification is allowed over time variables. 
As an example, ``the bounded response time'' is expressed in RTTL as follows \cite{Ost92}:

\begin{itemize}
\item $\square T [(red \wedge t=T) \rightarrow \lozenge (green \wedge T+3\leq t \leq T+5)]$
\end{itemize}

\noindent which means that ``if the traffic light is \emph{red} at time \emph{T}, then eventually within 3 to 5 ticks from \emph{T} the light must turn \emph{green}''. Above \emph{t} is the clock variable, and  \emph{T} is time variable, which is quantified in the formula. 

RTTL provides an explicit reference to clock value and indirect quantification to time values. This results in a very expressive language, and allows to write very complex quantitative constraints. This makes this logic very useful in real-time system specification. However, undecidability is a major problem. In addition, due to explicit clock reference, formulas become too complex and difficult to  understand. 

In addition to  discrete semantics, RTTL formulas can be also interpreted over dense time domains. The logic is undecidable in both discrete and dense semantics \cite{AH90}. The model checking in RTTL is also undecidable. RTTL has a sound proof system \cite{Ost89}.

Some decidable fragments of RTTL are presented in the literature. Some well-known fragments are as follows:

XCTL \cite{HLP90} is a propositional fragment of RTTL. It is an explicit clock logic, and it is interpreted over discrete time. XCTL has a less restricted quantification than RTTL in the sense that time variables can be quantified with only one outermost quantification; but the syntax of XCTL allows expressions with arithmetic operations. The satisfiability and model checking problems for XCTL with dense time semantics are both undecidable  \cite{HLP90}. However, these problems are PSPACE-complete for XCTL without quantification \cite{HLP90}. \cite{HLP90} provides a ``single exponent decision procedure for the validity of XCTL formulas'' and a ``double exponent procedure'' for XCTL model checking. 

TPTL \cite{AH90} is also a propositional fragment of RTTL, which is interpreted over discrete time. TPTL allows expressions with arithmetic operations; but this is only allowed for integer constants (not for variables). In TPTL explicit reference to clock is replaced by ``freezing'' quantification, and clock values are recorded through ``auxiliary static timing variables'' \cite{Ost92}.  The satisfiability and model checking problems for TPTL with discrete time semantics are EXPSPACE-complete; but they become undecidable with dense time semantics \cite{AH89}. \cite{AH89} presents a doubly-exponential-time decision procedure for TPTL. The model checking algorithm for the logic is ``exponential on the value of the product of all time constants'' \cite{Ost92}. \cite{AH90} shows that if past operators are added to the logic, the satisfiability problem for TPTL becomes non-elementary.  \cite{Hen91} proves that there is a complete finite axiomatization for TPTL with discrete time semantics.

\subsection{Metric Temporal Logic (MTL): }

MTL \cite{Koy90} is a propositional bounded-operator logic, which is a fragment of RTTL such that time references are added to temporal operators (`until', `next' and `since'). In MTL explicit reference to clock is not allowed, which makes the logic more practical because quantifications on a temporal domain are no longer needed. For example, the formula $A \rightarrow \lozenge_{\leq 10} B$ asserts that if \emph{A} occurs then \emph{B} occurs within 10 time units. 

MTL is interpreted over linearly ordered discrete time points.  In \cite{Koy90}, dense time domain is assumed. This allows MTL to express properties which cannot be precisely expressed in discrete-time domain, such as variables based on continuous time (e.g temperature and pressure) \cite{Ost92}. 

\cite{AH90} states that both the satisfiability and model checking problems for MTL over dense time domain are undecidable, but a deductive proof system exists. \cite{AH90} also shows that in case of discrete time they reduce to 
EXPSPACE-complete. \cite{AH90} also introduces a decision procedure for MTL over discrete time domain, which has 
2-EXPTIME complexity, and a model checking algorithm,  which is exponential on the value of the largest time constant. \cite{Koy90} provides a sound axiomatic system for MTL.  In \cite{HLN90} it is shown that XCTL and MTL cannot be compared; namely, for both logics, there is a property which is expressible in one logic, but not in the other \cite{Ost92}. However, in case of discrete time, ``TPTL and MTL are equally expressive (it is shown that this is not valid in dense domains \cite{Hen91})'' \cite{Ost92}. \cite{OW05} finds that ``the satisfiability problem for MTL over finite timed words is decidable, with non-primitive recursive complexity''.

In \cite{AFH91} MTL is restricted to ``interval-based strictly-monotonic real-time semantics''. This logic is called MITL, which uses operators with bound. In MITL point intervals are not allowed.  For example, the formula $\square (p \rightarrow \lozenge_{[3,3]} q)$ is not a valid formula because equality constraints are not allowed \cite{AHT92}. MITL cannot formalise punctuality properties\footnote{A \emph{punctuality} property states that the event \emph{B} follows \emph{A} in exactly $t$ seconds;  for any formal language that can express punctuality, the satisfiability problem is undecidable for a dense time domain \cite{Ost92}.}. Undecidability of logics interpreted over dense time is related to punctuality properties \cite{AHT92}. \cite{AFH91} shows that the satisfiability and model checking problems for MITL were shown to be EXPSPACE-complete. There is also a model checking algorithm, which is 2-EXPTIME. 

Recently, \cite{OW05,LW08} showed that restricting MTL to positive-length intervals is not necessary to achieve the decidability. They show that ``MTL over finitary event-based semantics'' are decidable without this restriction.  \cite{MNP05} compares the past and future fragments of  MITL with respect to the ``recognizability of their models by deterministic timed automata''. The authors show that ``timed languages specified by the past fragment of MITL, can be accepted by deterministic timed automata; but certain languages expressed in the future fragment of MITL are not deterministic.''

\subsection{Real-Time Interval Logic (RTIL): }

RTIL \cite{RG89} is a real-time interval logic with metric for time.  RTIL a propositional logic which allows to assign numerical values to interval bounds and to measure interval durations.  It also allows quantification over finite domains. Time points can be specified explicitly or relative to the beginning of the interval  \cite{BMN00}. These characteristics make RTIL to be useful in formalise specifications in a neater syntax. 

The specification in Section \ref{sub:Real-Time-Logic-(RTL):} is specified in RTIL as follows \cite{BMN00}:

\begin{itemize}
\item $\square\left[\odot B\hookrightarrow t_{b}\right]^{*}(\odot startA\Rightarrow\odot endA)$
\end{itemize}

\noindent where $\odot A$ extracts the time point at which \emph{A} becomes true, and the operator$*$ means there exists a subinterval.

\subsection{Temporal Interval Logic with Compositional Operators (TILCO):}

TILCO \cite{Mat96,MN96} is an extension of first-order logic with temporal operators, which do not have explicit temporal quantification. TILCO is an interval logic; that is, the logic is interpreted over linear intervals.   

TILCO can specify events and their relations (e.g. ordering, delay, etc.) in either qualitative or quantitative manner. Namely, end points of an interval at which an action or an event holds can be specified with respect to that of other events of actions; in addition, this can be done with an absolute numerical measure.  This makes TILCO a very expressive logic, and very useful to specify complex behaviours of real-time systems. 

Since TILCO is an interval-based logic, it is more natural to specify temporal constraints with time bounds.  Therefore, TILCO is very efficient to express ``invariants, precedence among events, periodicity, liveness and safety conditions, etc.'' \cite{BMN00}. 

The specification in Section \ref{sub:Real-Time-Logic-(RTL):} can be expressed in TILCO as follows \cite{BMN00}:

\begin{itemize}
\item $B\rightarrow endA?(0,t_{b})$ $\wedge$ $\neg\mathbf{until}(endA,\neg startA)$
\end{itemize}

\noindent where $?$ denotes universal temporal quantification.

 \cite{Mat96,MN00} provides a sound sound deductive system. This proof system is used along with the Isabelle theorem prover  \cite{Pau94} to provide an automatic proof tool for TILCO. The logic is unsurprisingly undecidable, because it extends the first-order logic. However, a decidable subset can be obtained if we restrict ourselves to quantifications on finite sets.
 
\section{Probabilistic Logics}

Probabilistic reasoning has been the subject of computer science for a long time. There is an extensive study about formal systems with uncertainty. There are two main approaches: extending classical logic with probabilistic operators (such as modal logic of knowledge in \cite{FH94}); combining probabilistic approach with non-classical logics (probabilistic extension of intuitionistic logic \cite{MOR03}). Below we review well-known probabilistic temporal logics.
 
\subsection{Probabilistic Temporal Logics}

\subsubsection{The Logics PCTL and PCTL*}

Probabilistic Computation Tree Logic, \emph{PCTL} \cite{HJ89,HJ94}, is a probabilistic extension of the branching time temporal logic CTL \cite{CES86}. PCTL is interpreted over discrete-time Markov chains. Each transition in a path corresponds to one time step. The path quantifiers in classical branching-time temporal logics are replaced with probabilities. Namely, universal and existential quantification over paths is a subset of probabilistic quantification. PCTL's probabilistic operator provides a more general quantification, because as well as expressing a property is true at all/some paths, we can also express a property is true at more than 50\% of the paths. 

PCTL is very convenient to specify so-called \emph{soft deadline properties}, e.g. ``after a request for a service, there is at least a 98\% probability that the service will be carried out within 2 seconds'' \cite{HJ94}. Soft deadline properties are important in real-time system specification. 

Some real-time requirements are specified in PCTL as follows \cite{HJ94}:

\begin{itemize}
\item \emph{(i)} $\forall \Box f \equiv f\mathcal{U}_{\geq 1}^{\leq \infty} false$  $\qquad$ \emph{(ii)} $\exists \diamondsuit f \equiv true U_{> 0}^{\leq \infty} f$. 
\end{itemize}

\noindent where $f_{1} \mathcal{U}_{\geq t}^{\leq p} f_{2}$ asserts that ``there is at least a probability \emph{p} that either $f_{1}$ will remain true for at least \emph{t} time units, or that both $f_{2}$ will become true within \emph{t} time units and that $f_{1}$ will be true from now on until $f_{2}$ becomes true''; and $f_{1} U_{\geq t}^{\leq p} f_{2}$ asserts that ``there is at least a probability \emph{p} that both $f_{2}$ will become true within \emph{t} time units and that $f_{1}$ will be true from now on until $f_{2}$ becomes true'' \cite{HJ94}. Therefore, ``$\forall \Box f$ intuitively means that $f$ is always true (in all states that can be reached with non-zero probability)'', and ``$\exists \diamondsuit f$  means that there exists a state where $f$ holds which can be reached with non-zero probability'' \cite{HJ94}.

\cite{HJ94} presents a model checking algorithm for  PCTL, which is polynomially bounded by the size of the formula  and the \emph{Markov chain}\footnote{A \emph{Markov chain} is a tuple $(S,P)$ where \emph{S} is a set of \emph{states} and $P: S \times S \rightarrow$ [0,1] is the \emph{transition probability matrix} such that  $(\forall \emph{s} \in \emph{S}) \sum_{s' \in S}P(s,s') = 1$ \cite{Rab03}.} model.

\cite{ASB95} defines another probabilistic variant of CTL \cite{CES86}. This new logic is called \emph{PCTL*}, which can specify quantitative probabilistic properties of systems, modelled as discrete \emph{Markov processes}\footnote{A (finite) \emph{Markov process} is a 4-tuple $(AP,S,P,\mathcal{L})$, where $AP$ is a finite set of \emph{atomic propositions}, $S$ is a countable nonempty set of \emph{states}, $P: S \times S \rightarrow$ [0,1] is the \emph{transition probability function} such that $(\forall \emph{s} \in \emph{S}) \sum_{s' \in S}P(s,s') = 1$ and $\mathcal{L} : S \rightarrow 2^{AP}$ is a \emph{labeling function} \cite{ASB95}.}. \cite{CES86} also extends discrete Markov processes to  \emph{generalized Markov processes}\footnote{A \emph{generalized Markov process} is a 3-tuple $(AP,S,\mathcal{L})$ (where $AP,S$ and $\mathcal{L}$ are defined as in Markov processes) and a finite set of constraints on the transition probabilities \cite{ASB95}.}, where transition probability function is not total. Generalized Markov processes are convenient to model `'abstraction'' and ``refinement''.  \cite{ASB95} also presents an elementary model checking algorithm for PCTL* over discrete Markov processes, which is then extended for generalized discrete Markov processes. This algorithm can also be used to determine the satisfiability of PCTL* formulas. In fact, \cite{ASB95} shows that the decision problem for PCTL* formulas on generalized Markov processes is decidable. However, no efficient computational method is given for this problem. In addition, no sound and complete axiomatisation of the logic is given.

\cite{BA95} shows that model-checking algorithms for extensions of PCTL and PCTL* to probabilistic-nondeterministic models have a polynomial-time complexity in the size of the model, which is same as the model checking complexity on Markov chains  \cite{HJ89,HJ94,ASB95}. This result shows that adding nondeterminism does not increase model checking complexity in the size of the model. When we consider time bounds expressed in terms of the size of the formula, the situation is different. The model checking complexity of PCTL is linearly bounded in the size of the formula for both Markov chains and probabilistic-nondeterministic systems. However, while model checking complexity of PCTL* on Markov chain is exponentially bounded in the size of formula, it is in double exponential time on probabilistic-nondeterministic systems. 

\subsubsection{The Logic PTCTL}

A probabilistic extension of the real-time branching logic TCTL is defined in \cite{KNR99}. The logic is called \emph{PTCTL}, which combines both the logics TCTL and PCTL. PTCTL can formalize properties such as `with a probability higher than 0.9 the message is delivered within 5 seconds'. This can be expressed in PTCTL as follows: $\textup{P}_{> 0.9} [true \; \textup{U}^{\leq 5} \; rcv]$.  PTCTL includes a set of clock variables in its syntax in order to specify timing properties. 

Since PTCTL is a probabilistic extension of TCTL, PTCTL is also an undecidable logic. \cite{KNR99} shows that the model checking problem is ``polynomial in the size of region graph and linear in the size of formula''. It follows that the model checking problem is EXPTIME due to the size of region graph. \cite{JLS07} shows that the model checking problem is EXPTIME-complete. \cite{JLS07} also considers the model checking problems of some subclasses of PTCTL.  

\subsubsection{The Logic PLTL}
A propositional probabilistic discrete-linear temporal logic, called Probabilistic Propositional Temporal Logic (\textit{PLTL}), is introduced in \cite{Ogn06}. PLTL allows probabilistic reasoning, which is extended with temporal aspects. The logic is interpreted over linear time points, and includes standard temporal operators, such as `next', `until', `sometime' and `always'. PLTL can express sentences such as ``(according to the current set of information) the probability that sometime in the future $\alpha$ is true is at least \textit{n}''  \cite{Ogn06}.

Given that $\bigcirc$, $\diamondsuit$ and $\Box$ are the `next', `sometime' and `always' operators, respectively, and $P_{\sim \alpha}$ ($\sim \in \{ <,\leq,=,\geq,> \}$) is a probabilistic operator, an example of a PLTL formula is  \cite{Ogn06}

\begin{itemize}
\item $\bigcirc P_{\geq r}p \wedge \diamondsuit P_{< s}(p \rightarrow q) \rightarrow \Box P_{=t} q$ \end{itemize}

\noindent which aserts ``if the probability of \textit{p} in the next moment is at least \textit{r} and sometime in the future \textit{q} follows from \textit{p} with the probability less than \textit{s}, then the probability of \textit{q} will always be equal to \textit{t}''  \cite{Ogn06}.

\cite{Ogn06} analyses completeness, decidability and complexity of the logic PLTL. It describes a class of so-called `measurable models'. It is proved that ``PLTL restricted to the class of all measurable models ($PLTL_{Meas}$)'' has a sound and complete (infinitary) axiomatisation. The term infinitary means that the language and formulas are finite, but proofs can be infinite (The completeness cannot be proved with finitary axiomatisation). \cite{Ogn06}  shows that ``a $PLTL_{Meas}$-satisfiable formula is satisfiable in an ultimately periodic model in which various parameters are bounded by functions depending on the size of the formula''. \cite{Ogn06} also shows that ``the satisfiability problem for  $PLTL_{Meas}$ is PSPACE-hard, and that it belongs to NEXPTIME''.

In \cite{Ogn06} also introduces First-order Probabilistic Temporal Logic (\textit{FOPLTL}), which is the first-order version of PLTL. The complete infinitary axiomatisation is extended for the logic FOPLTL (No complete finitary axiomatisation is possible). The set of all FOPLTL-valid sentences is not recursively enumerable \cite{GHR94}.

\subsubsection{The Logics $PTL_{f}$ and $PTL_{b}$}

\cite{HS84} introduces two probabilistic branching time temporal logics $PTL_{f}$  and $PTL_{b}$, which are interpreted over finite Markov chains and stochastic processes, respectively. $PTL_{f}$ and $PTL_{b}$ can express properties, such as 
``invariant and liveness without explicit reference to the values of the transition probabilities'' \cite{HS84}. $PTL_{f}$ is a suitable logic for the specification of sequential programs. $PTL_{b}$ is an extension of $PTL_{f}$, which can be used to reason about concurrent programs. 

To show the syntax of the logics, let us consider the formula $p \; \forall U q$. This formula asserts that ``along all paths \emph{w} starting with the initial state and consisting only of transitions with nonzero probability, \emph{p} holds at all states of \emph{w} up to the first state, if any, at which \emph{q} holds'' \cite{HS84}.

The satisfiability problems of $PTL_{f}$  and $PTL_{b}$ are decidable. \cite{HS84} provides an EXPTIME decision procedure based on the tableau techniques of \cite{BMP81} and \cite{CE82}. \cite{HS84}  provides proof systems for both logics. The paper also shows that $PTL_{b}$ does not have a finite-model property, and there is a connection between ``satisfiable formulas of $PTL_{b}$ and finite state concurrent probabilistic programs''.

In literature, we can find similar formal systems to $PTL_{f}$  and $PTL_{b}$. \cite{Pnu83} proposes a linear time probabilistic logic to reason about concurrent probabilistic programs; but it is not a complete logic.  \cite{LS83} introduces a similar logic which is more expressive than $PTL_{b}$; but its decision procedure is less efficient. \cite{CY88} determines ``the complexity of testing whether a finite state (sequential or concurrent) probabilistic program satisfies its specification expressed in linear temporal logic LTL''. \cite{CY88} shows that this problem is decidable and it is in PSPACE. \cite{CY88} also provides an EXPTIME procedure for sequential programs. This is a more efficient method than that of $PTL_{f}$ and $PTL_{b}$. For concurrent programs it is shown that the problem is complete in 2-EXPTIME.

\subsubsection{The Logic PDC}

The Probabilistic Duration Calculus (\textit{PDC})  \cite{LRSZ92} is an extension of Duration Calculus \cite{CHR91} with probabilities. PDC allows us to reason about probabilistic systems, and enables to express requirements such as a property holds with a certain probability. In PDC the system model is described as a finite automaton with fixed transition probabilities, which actually defines a discrete Markov process. The main idea is described in \cite{LRSZ92} is to express properties in DC, define satisfaction probabilities for formulas, and define a calculus to calculate the probability of a formula from its  subformulas' probabilities.  

PDC satisfiability is described in \cite{LRSZ92} as follows: ``Consider some finite probabilistic timed automata \textit{A}. The behaviours of \textit{A} can be represented as a set of \textit{M} of DC models. The probabilistic principles that manage the working of \textit{A} used to introduce probability on the subsets of \textit{M}. Given a DC formula \textit{D}, the term $\pi(D)(t)$ denotes the probability of those models from \textit{M} that satisfy \textit{D} at the interval $[0,t]$. A term of this sort is the component of PDC language'' \cite{LRSZ92}. An example PDC formulas is given below:

\begin{itemize}
\item $\pi_{s_{0}}((true;\lceil s \rceil);(\lceil s' \rceil;true))(t)=0$
\end{itemize}

In \cite{LRSZ92} PDC is interpreted over discrete time; i.e. discrete transitions are assumed in models, defined as probabilistic time automata. In a later work, \cite{HC94}, PDC was defined for the case of continuous time, in which  transitions in probabilistic automata take place in continuous time. In this logic, properties are written in terms of  DC formulas. ``Implementations of given requirements are modelled by continuous semi-Markov processes with finite space, which are expressed as finite automata with stochastic delays of state transitions (such an automaton is called continuous time probabilistic automaton)'' \cite{HC94}. \cite{HC94} also defines a probabilistic model for DC formulas and a set of axioms/rules to calculate the satisfaction probabilities of DC formulas with respect to probabilistic automata. To our best knowledge, there is no complete proof system for PDC. As for the decidability,  PDC is, not surprisingly, an undecidable logic.

\cite{HZ07} defines the logic \emph{Simple Probabilistic Duration Calculus (SPDC)}, which is another probabilistic extension of Duration Calculus. The syntax of SPDC allows us ``to reason about the probability of the satisfaction of a duration formula by a probabilistic timed automaton as well as to specify real-time properties of the system itself''. SPDC is interpreted over \emph{behavioural models}\footnote{A \emph{behavioural model} is a variant of probabilistic timed automata, where probabilistic transitions are discrete. ``To resolve the nondeterminism between the passage of time and discrete transitions they use the concept of \emph{adversary} which is essentially a deterministic schedule policy. Then, the set of executions of a probabilistic time automaton according to an adversary forms a Markov chain, and hence the satisfaction of a probabilistic CTL formula by this set can be defined, and then based on the region graph of the timed automaton the satisfaction of a probabilistic CTL formula by the timed automaton can be also verified'' \cite{CH06}.}, proposed in \cite{KNSS02}, which are variant of probabilistic timed automata. \cite{HZ07} proposes a model checking technique which is an extension of the technique introduced in \cite{TH04} ``to check if a timed automaton satisfies a DC formula in the form of \emph{linear duration invariants} or discretisable DC formulas based on searching the integral reachability graph of the timed automaton'' \cite{HZ07} . The model checking problem is decidable ``for a class of SPDC formulas of the form \emph{linear duration invariants}, or a formula for bounded liveness'' \cite{HZ07}.

\subsubsection{The Logic PNL}

\cite{Gue00b} introduces the Probabilistic Neighbourhood Logic (\textit{PNL}), which extends Neighbourhood Logic.
\cite{Gue00b} provides a complete proof system by extending the proof system of NL. In PNL, a more generalised version of probabilistic timed automata (defined in \cite{HC94}) is assumed. 

PNL has a similar grammar to the logic NL. It contains duration operators and probabilistic operators. The function symbols take a duration as argument and return a term of the probability. We now consider an example. Let \emph{b} denote a formula which is true at any interval between two consecutive processes. The following formula expresses ``the assumption that the probability for the duration of such a period to be no bigger than \emph{x} is a function of \emph{x} which is the interpretation of the function symbol \emph{F} in the model''  \cite{Gue00c}:

\begin{itemize}
\item $p(b\wedge\ell\leq x = F(x))$
\end{itemize}

PNL has the same expressive power with PDC, except for state expressions and their durations. Since PNL is an extension of NL, it is an undecidable logic.

\subsection{Probabilistic Dynamic Logics}

Since Kozen's definition of formal semantics of probabilistic programs  \cite{Koz79}, some work has been done in this direction. Several systems have been introduced to formally study probabilistic programs. In particular, \emph{probabilistic dynamic logics} received considerable attention. Some historical development in this area is given below:

In \cite{FH82}, Feldman and Harel introduced a first-order probabilistic dynamic logic, called \emph{Pr(DL)}, which can express properties of probabilistic programs. The syntax of this logic is similar to that of Pratt's first-order dynamic logic \cite{Pra76}. The semantics of Pr(DL) is based on extension of Kozen's formal semantics of probabilistic programs \cite{Koz79}. \cite{FH82} provides a complete proof system for Pr(DL) relative to first-order analysis. \cite{FH82} shows that for discrete probabilities the logic reduces to first-order analysis with integer variables.  Since the underlying theory is highly undecidable, the logic  Pr(DL) is also undecidable.

On propositional level, the well-known logics are Feldman's  \emph{P-Pr(DL)} \cite{Fei83} and Kozen's \emph{PPDL} \cite{Koz83}. \cite{Fei83} defines the logic P-Pr(DL), which is a propositional fragment of the first-order dynamic logic Pr(DL). P-Pr(DL) has many important characteristics of Pr(DL), such as ``the ability to use full first-order real-number theory for dealing with probabilities, and deterministic regular programs, while still being decidable'' \cite{Fei83}. Neither the complexity of the decision procedure, nor a proof system is provided. 

In \cite{Koz83} a probabilistic analog \emph{PPDL} of Propositional Dynamic Logic is introduced. \cite{Koz83} proves the  finite model property by showing that models can be reduced to an equivalent finite model with a bound on the number of states. A polynomial-space decision procedure is given to decide the validity of programs. \cite{Koz83} also provides ``a deductive calculus'' and shows its usefulness on an example program.

In \cite{Fel84} a Propositional Dynamic Logic with \emph{explicit} probabilities is introduced. The language allows formulas of propositional probabilistic programs, where probabilistic operators are applied in a limited form.  \cite{Fel84} provides a 2-EXSPACE decision procedure for the logic by reducing it to ``the decision problem of the theory of real closed fields''. 

\cite{Seg71} introduces a family of propositional calculi of qualitative probabilities (\emph{QP}) with one binary operator $\leq$. $\leq$ intuitively means ``at least as probable as''. Given that $\varphi$ and $\psi$ are two arbitrary QP formulas, $\varphi\leq\psi$ means that ``the probability of $\varphi$ is not greater than the probability of $\psi$'' \cite{Gue99}.  \cite{Seg71} presents a complete deductive system for QP, and shows that QP is decidable.

\cite{Gue99} extends QP with ``many $\leq$-operators and operations among them that are analogous to the operations of composition, union, and iteration on modal operators known in propositional dynamic logic''.  The resulting logic (\emph{DQP}) allows us to reason about probabilistic processes. The formula $w\models\varphi\leq_{t}\psi$ intuitively means that ``the probability for a transition (experiment) \emph{t} to transform \emph{w} into a possible world that satisfies $\varphi$ is smaller or equal to the probability for \emph{t} to transform \emph{w} into a possible world that satisfies $\psi$''  \cite{Gue99}.  An $\omega$-complete proof system is presented for DQP in \cite{Gue99}, which requires to build an infinite canonical model. This implies that DQP is undecidable. 

\subsection{Probabilistic Mu-Calculus}

\cite{CIN05} presents the logic \emph{Generalised Probabilistic Logic (GPL)}, which is a Mu-Calculus-based modal logic, in order to reason about ``reactive probabilistic labelled transition systems (RPLTSs)''. An RPLTS structure includes (probabilistic) transitions and (nonprobabilistic) actions, where nonprobabilistic actions are chosen \emph{externally}, in contrary to Markov decision processes where nonprobabilistic choices are done \emph{internally}. 

To show the syntax of GPL, we consider the following example: $P_{\geq 1}(\nu X.\phi \wedge [.][.]X)$ informally means that ``it is almost always true that $\phi$ holds at all even time instants'' ($[.]\phi$ $\equiv$ $\bigwedge_{a \in Act}[a]\phi$, where \emph{Act} denotes a set of actions) \cite{CIN05}.

GPL can be considered as a framework to define temporal logics on reactive models. GPL is an expressive logic. Some standard probabilistic (modal/temporal) logics, such as PCTL*, are subsumed by the logic GPL. \cite{CIN05} presents a model-checking algorithm which employs techniques to solve non-linear equations.

\subsection{Probabilistic Instuitionistic Logics}

A probabilistic extension of propositional instuitionistic logic is introduced in \cite{MOR03}, where a view of instuitionistic logic is described as ``in addition to propositions which are proved to be true and those which are proved to be false, there is a third class of propositions which may turn out either way and intuitionism allows us to reason about them''. The propositional instuitionistic language is enriched with probabilistic operator, resulting in the operator $P_{\geq n}\alpha$, which informally means that  ``the probability of truthfulness of $\alpha$ is at least \textit{n}'' \cite{MOR03}. The logic does not allow nested probabilistic operators. Probabilistic instuitionistic logic is interpreted over a combined model of instuitionistic Kripke models and probabilities. \cite{MOR03} proves that the logic is decidable, and presents a sound and complete proof system. 

\subsection{Probabilistic Logics with New Types of Probability Operators}

\cite{OR99} introduces a family of probabilistic logics, called $LP_{P,Q,O}$,  with new types of probabilistic operators, which have the form $Q_{F}$, where ``\emph{F} is a set from a recursive family \emph{O} of recursive rational subsets of $[0,1]$''. $Q_{F}\alpha$ states that the probability of $\alpha$ is within the set \emph{F}. The authors assume the so called \emph{measurable} models, different from probabilistic models based on Kripke structures.  

$LP_{P,Q,O}$'s unique operator $Q_{F}$ can specify richer probabilistic expressions, which cannot be expressed by standard probabilistic logics, such as PCTL, because operator $Q_{F}$ cannot be translated into $P_{\geq}$-like operators. For example, assume the model describes tossing a coin finitely many times. Given that $\alpha$ means that `it comes up heads', and $F = \{\frac{1}{2},\frac{1}{2^{2}},\frac{1}{2^{3}},...\}$ \cite{OR99}. Clearly, $Q_{F}\alpha$ is true in the model. However, $Q_{F}\alpha$ cannot be expressed in classical probabilistic logics, such as PCTL, because $Q_{F}$ cannot be translated into $P_{\geq}$-like operators. 

The choice of the family \emph{O} of recursive rational subsets of $[0,1]$ appearing in \emph{Q} is very important, because this choice determines the language of the logic. The choice is also important for the decidability and expressiveness of  the resulting logic. Although the logic $LP_{P,Q,O}$ is not decidable in general, \cite{OR99}  provides a sublanguage which is shown to be decidable. \cite{OR99} also provides a sound and complete axiomatic systems for $LP_{P,Q,O}$.

\subsection{Probabilistic Logics for Reasoning About Knowledge and Uncertainty}


Halpern et. al., in a series of articles, studied reasoning about knowledge and probability.  In \cite{FHM90} a language is defined which can express statements such as ``the probability of $E_{1}$ is less than 1/3 and the probability of $E_{1}$ is at least twice the probability of $E_{2}$'', where $E_{1}$ and $E_{2}$ are events. \cite{FHM90} considers both the situations where all events are \emph{measurable} or  all events are \emph{nonmeasurable} (i.e. events that are not assigned a probability). A sound and complete proof system is presented for both \emph{measurable} and \emph{nonmeasurable} cases. The satisfiability problem is NP-complete for both cases.


Some related works to that of \cite{FHM90} is as follows: \cite{GKP88} presents a less expressive logic, which is shown to be NP-complete. The measurable case of the logic proposed by \cite{FHM90} can be considered as a fragment of the Probabilistic PDL by \cite{Fel84}. \cite{Koz83} also considers a Probabilistic PDL, which is PSPACE-complete; but this logic is not closed under Boolean combination, and it does not allow nonlinear combinations.

\cite{FH91} introduces a new approach to deal with uncertainty. Namely, it does not require assigning a probability value to every event. For nonmeasurable events the paper considers the \emph{inner} measure and \emph{outer} measure of events. The paper states that  ``inner measures induced by probability measures turn out to correspond in a precise sense to Dempster-Shafer belief functions \cite{Sha76}; hence, in addition to providing promising new conceptual tools for dealing with uncertainty, this approach shows that a key part of the important Dempster-Shafer theory of evidence is firmly rooted in classical probability theory.''

\cite{FH94} presents a probabilistic logic which is an extension of the logic defined in \cite{FHM90} (which is itself a formalisation of Nilsson's probability logic \cite{Nil86}). Indeed, the logic of \cite{FH94} is a probabilistic extension of the logic of knowledge, which can express the statements  such as ``according to agent \emph{t}, formula $\varphi$ holds with probability at least \emph{b}'' \cite{FHM90}. The language allows one to compare the probabilities of events for each agent. \cite{FH94} provides a complete proof system. It proves the decidability through some decision procedures. \cite{FH94} also considers the extended language with ``common knowledge and a probabilistic variant of common knowledge''.

 \cite{AH94} analyses decidability and expressiveness of probabilistic first-order logics. It is shown that for discrete probabilities such logics are undecidable.  If arbitrary probability distributions are assumed, the situation becomes even worse. Not surprisingly, sound and complete proof systems are not available for such logics. \cite{AH94} shows that for the following cases complete axiomatic systems can be found: ``the language consists only of unary predicates and the case where we restrict to bounded domains; in particular, when combined with the standard axioms for reasoning about first-order logic, the axioms for reasoning about probabilities over the domain are complete for a language if it contains only unary predicates; when combined with axioms for equality and an axiom that says that the domain has at most \emph{n} elements, the axioms are complete for the language if we restrict attention to domains with at most \emph{n} elements.''

\section{Conclusion} \label{sec:conclusion}

In this paper we have analysed well-known real-time temporal logics and probabilistic temporal logics. We extrapolated the notions of decidability, axiomatizability, expressiveness, model checking, etc. for each logic analysed, whenever possible. For a comparison of features of the temporal logics we discussed see Table 1. Note that we use the following abbreviations: \emph{No*}: Undecidable in general, but decidable for some fragments or specific cases; \emph{No**}: No deduction system in general, but available for some fragments or specific cases; \emph{No***}: No model checking algorithm in general, but available for some fragments or specific cases; \emph{Yes*}: Decidable for some time domains; \emph{Yes**}:  Available for some time domains; \emph{Yes***}: Available for some time domains.

\begin{landscape}
\begin{table}
\caption{A comparison of features of temporal logics.}
\centering
\small
\begin{tabular}{|c|c|c|c|c|c|c|c|}
\hline
\textbf{Logic} & \textbf{Logic Order} & \textbf{Fund. Entity} & \textbf{Temp. Struc.} &\textbf{Metric for Time} & \textbf{Decidability} & \textbf{Deductive Sys.} & \textbf{Model Checking}\tabularnewline\hline\hline
TCTL&Propositional&Point&Branching&Yes&No&?&Yes\tabularnewline\hline
RTCTL&Propositional&Point&Branching&Yes&Yes&?&Yes\tabularnewline\hline
TPCTL&Propositional&Point&Branching&Yes&Yes&?&Yes\tabularnewline\hline\hline
RTL&First-order&Interval&Linear&Yes&No*&No&No***\tabularnewline\hline
RTIL&Propositional&Interval&Linear&Yes&Yes&No&?\tabularnewline\hline
RTTL&First-order&Point&Linear&Yes&No&Yes&No\tabularnewline\hline
TPTL&Propositional&Point&Linear&Yes&Yes*&Yes**&Yes***\tabularnewline\hline
MTL&Propositional&Point&Linear&Yes&Yes*&Yes&Yes***\tabularnewline\hline
MTIL&Propositional&Interval&Linear&Yes&Yes&?&Yes\tabularnewline\hline
XCTL&Propositional&Point&?&Yes&Yes*&?&Yes***\tabularnewline\hline
TILCO&First-order&Interval&Linear&Yes&No*&Yes&No***\tabularnewline\hline\hline
PCTL&Propositional&Point&Branching&No&Yes&?&Yes\tabularnewline\hline
PCTL*&Propositional&Point&Branching&No&Yes&?&Yes\tabularnewline\hline
PLTL&Propositional&Point&Linear&No&No*&No**&No\tabularnewline\hline
$PTL_{f}$&Propositional&Point&Branching&No&Yes&Yes&?\tabularnewline\hline
$PTL_{b}$&Propositional&Point&Branching&No&Yes&Yes&?\tabularnewline\hline
PDC&First-order&Interval&Linear&Yes&No&?&?\tabularnewline\hline
PNL&First-order&Interval&Linear&Yes&No&Yes&?\tabularnewline\hline\hline
\end{tabular}
\end{table}
\end{landscape}

\subsection*{Acknowledgements.}
This work was partially supported by EPSRC research project EP/F033567.

\bibliographystyle{plain}       
\bibliography{bibliography}     

\end{document}